\def\preflag{y}
\nopagenumbers
%
%
%
\def\unredoffs{} \def\redoffs{\voffset=-.31truein\hoffset=-.59truein}
\def\speclscape{}
%
%
%
%
\magnification=1200
\hsize=15.1truecm
\vsize=23.5truecm 
\def\prepr{y}
\def\header{\global\voffset=1truecm\advance\vsize by-\voffset
            \headline=
            {\ifnum\pageno>1 
               \ifx\prepr\preflag
                 \vbox{\hsize=15truecm\advance\hsize by -2pt\hfill}
               \else
                 \ifodd\pageno
                   \vbox{\hsize=15truecm\advance\hsize by -2pt
                    \centerline{\runtitle}}
                 \else
                   \vbox{\hsize=15truecm\advance\hsize by -2pt
                    \centerline{\runauth}}
                 \fi
               \fi
             \else 
               \ifx\preflag\prepr
                 \vbox{\hsize=15.0truecm
\centerline{\hbox{\vbox{\baselineskip 0.20truein\noindent\anbf To 
Appear in the {\anbfsl Proceedings of the
VIII$^{\hbox{\anbfsls th}}$ Topical Meeting on Optics of Liquid Crystals}, 
27 September-1 October 1999,
Humacao, PR, USA ({\anbfsl Mol. Cryst. Liq. Cryst.})}}}}
               \fi
             \fi}}
\newbox\leftpage \newdimen\fullhsize \newdimen\hstitle \newdimen\hsbody
\tolerance=1000\hfuzz=2pt
\catcode`\@=11 
\def\bigans{b }
\def\answ{b }
%

\ifx\answ\bigans\message{(This will come out unreduced.}
\unredoffs\baselineskip=.24truein plus 2pt minus 1pt
\hsbody=\hsize \hstitle=\hsize 
\else\message{(This will be reduced.} \let\l@r=L
\magnification=1000\baselineskip=16pt plus 2pt minus 1pt \vsize=7truein
\redoffs \hstitle=8truein\hsbody=4.75truein\fullhsize=10truein\hsize=\hsbody
\output={\ifnum\pageno=0 
  \shipout\vbox{\speclscape{\hsize\fullhsize\makeheadline}
   \hbox to \fullhsize{\hfill\pagebody\hfill}}\advancepageno
  \else
 \almostshipout{\leftline{\vbox{\pagebody\makefootline}}}\advancepageno 
  \fi}
\def\almostshipout#1{\if L\l@r \count1=1 \message{[\the\count0.\the\count1]}
      \global\setbox\leftpage=#1 \global\let\l@r=R
 \else \count1=2
  \shipout\vbox{\speclscape{\hsize\fullhsize\makeheadline}
      \hbox to\fullhsize{\box\leftpage\hfil#1}}  \global\let\l@r=L\fi}
\fi
%
\newcount\yearltd\yearltd=\year\advance\yearltd by -1900

\ifx\preflag\prepr\footline={\hss\tenrm\folio\hss}\else\footline={\hss}\fi
%

\def\draftmode{\message{ DRAFTMODE }\def\draftdate{{\rm preliminary draft:
\number\month/\number\day/\number\yearltd\ \ \hourmin}}%
\headline={\hfil\draftdate}\writelabels\baselineskip=20pt plus 2pt minus 2pt
 {\count255=\time\divide\count255 by 60 \xdef\hourmin{\number\count255}
  \multiply\count255 by-60\advance\count255 by\time
  \xdef\hourmin{\hourmin:\ifnum\count255<10 0\fi\the\count255}}}
\def\nolabels{\def\wrlabeL##1{}\def\eqlabeL##1{}\def\reflabeL##1{}}
\def\writelabels{\def\wrlabeL##1{\leavevmode\vadjust{\rlap{\smash%
{\line{{\escapechar=` \hfill\rlap{\sevenrm\hskip.03in\string##1}}}}}}}%
\def\eqlabeL##1{{\escapechar-1\rlap{\sevenrm\hskip.05in\string##1}}}%
\def\reflabeL##1{\noexpand\llap{\noexpand\sevenrm\string\string\string##1}}}
\nolabels
%
\global\newcount\secno \global\secno=0
\global\newcount\meqno \global\meqno=1
\def\newsec#1{\global\advance\secno by1\message{(\the\secno. #1)}

\global\subsecno=0\eqnres@t\noindent{\the\secno. #1}
\writetoca{{\secsym} {#1}}\par\nobreak\medskip\nobreak}
\def\eqnres@t{\xdef\secsym{\the\secno.}\global\meqno=1\bigbreak\bigskip}
\def\sequentialequations{\def\eqnres@t{\bigbreak}}\xdef\secsym{}
\global\newcount\subsecno \global\subsecno=0
\def\subsec#1{\global\advance\subsecno by1\message{(\secsym\the\subsecno. #1)}
\ifnum\lastpenalty>9000\else\bigbreak\fi
\noindent{\sl\secsym\the\subsecno. #1}\writetoca{\string\quad 
{\secsym\the\subsecno.} {#1}}\par\nobreak\medskip\nobreak}

\def\appendix#1#2{\global\meqno=1\global\subsecno=0\xdef\secsym{\hbox{#1.}}
\bigbreak\bigskip\noindent{\bf Appendix #1. #2}\message{(#1. #2)}
\writetoca{Appendix {#1.} {#2}}\par\nobreak\medskip\nobreak}
%
%
\def\eqnn#1{\xdef #1{(\secsym\the\meqno)}\writedef{#1\leftbracket#1}%
\global\advance\meqno by1\wrlabeL#1}
\def\eqna#1{\xdef #1##1{\hbox{$(\secsym\the\meqno##1)$}}
\writedef{#1\numbersign1\leftbracket#1{\numbersign1}}%
\global\advance\meqno by1\wrlabeL{#1$\{\}$}}
\def\eqn#1#2{\xdef #1{(\secsym\the\meqno)}\writedef{#1\leftbracket#1}%
\global\advance\meqno by1$$#2\eqno#1\eqlabeL#1$$}
%
\newskip\footskip\footskip14pt plus 1pt minus 1pt 
\def\footnotefont{\ninepoint}\def\f@t#1{\footnotefont #1\@foot}
\def\f@@t{\baselineskip\footskip\bgroup\footnotefont\aftergroup\@foot\let\next}
\setbox\strutbox=\hbox{\vrule height9.5pt depth4.5pt width0pt}
\global\newcount\ftno \global\ftno=0
\def\foot{\global\advance\ftno by1\footnote{$^{\the\ftno}$}}
%
\newwrite\ftfile   
\def\footend{\def\foot{\global\advance\ftno by1\chardef\wfile=\ftfile
$^{\the\ftno}$\ifnum\ftno=1\immediate\openout\ftfile=foots.tmp\fi%
\immediate\write\ftfile{\noexpand\smallskip%
\noexpand\item{f\the\ftno:\ }\pctsign}\findarg}%
\def\footatend{\vfill\eject\immediate\closeout\ftfile{\parindent=20pt
\centerline{\bf Footnotes}\nobreak\bigskip\input foots.tmp }}}
\def\footatend{}
%
%
\global\newcount\refno \global\refno=1
\newwrite\rfile
\def\ref{$^{\the\refno}$\nref}
\def\nref#1{\xdef#1{$^{\the\refno}$}\writedef{#1\leftbracket#1}%
\ifnum\refno=1\immediate\openout\rfile=refs.tmp\fi
\chardef\wfile=\rfile\immediate
\write\rfile{\noexpand\item{{\the\refno}.\ }\reflabeL{#1\hskip.31in}\pctsign}\global\advance\refno by1\findarg}
\def\findarg#1#{\begingroup\obeylines\newlinechar=`\^^M\pass@rg}
{\obeylines\gdef\pass@rg#1{\writ@line\relax #1^^M\hbox{}^^M}%
\gdef\writ@line#1^^M{\expandafter\toks0\expandafter{\striprel@x #1}%
\edef\next{\the\toks0}\ifx\next\em@rk\let\next=\endgroup\else\ifx\next\empty%
\else\immediate\write\wfile{\the\toks0}\fi\let\next=\writ@line\fi\next\relax}}
\def\striprel@x#1{} \def\em@rk{\hbox{}} 
\def\lref{\begingroup\obeylines\lr@f}
\def\lr@f#1#2{\gdef#1{\ref#1{#2}}\endgroup\unskip}

\def\addref#1{\immediate\write\rfile{\noexpand\item{}#1}} 
\def\startrefs#1{\immediate\openout\rfile=refs.tmp\refno=#1}
\def\xref{\expandafter\xr@f}\def\xr@f[#1]{#1}
\def\refs#1{\count255=1$^{\r@fs #1{\hbox{}}}$}
\def\r@fs#1{\ifx\und@fined#1\message{reflabel \string#1 is undefined.}%
\nref#1{need to supply reference \string#1.}\fi%
\vphantom{\hphantom{#1}}\edef\next{#1}\ifx\next\em@rk\def\next{}%
\else\ifx\next#1\ifodd\count255\relax\xref#1\count255=0\fi%
\else#1\count255=1\fi\let\next=\r@fs\fi\next}
%

%
\newwrite\ffile\global\newcount\figno \global\figno=1
\def\fig{Figure~\the\figno\nfig}
\def\nfig#1{\xdef#1{Figure~\the\figno}%
\writedef{#1\leftbracket fig.\noexpand~\the\figno}%
\ifnum\figno=1\immediate\openout\ffile=figs.tmp\fi\chardef\wfile=\ffile%
\immediate\write\ffile{\noexpand\medskip\noexpand\item{Fig.\ \the\figno. }
\reflabeL{#1\hskip.55in}\pctsign}\global\advance\figno by1\findarg}
\def\xfig{\expandafter\xf@g}\def\xf@g fig.\penalty\@M\ {}
\def\figs#1{figs.~\f@gs #1{\hbox{}}}
\def\f@gs#1{\edef\next{#1}\ifx\next\em@rk\def\next{}\else
\ifx\next#1\xfig #1\else#1\fi\let\next=\f@gs\fi\next}
\newwrite\lfile
{\escapechar-1\xdef\pctsign{\string\%}\xdef\leftbracket{\string\{}
\xdef\rightbracket{\string\}}\xdef\numbersign{\string\#}}

\def\writestop{\def\writestoppt{\immediate\write\lfile{\string\pageno%
\the\pageno\string\startrefs\leftbracket\the\refno\rightbracket%
\string\def\string\secsym\leftbracket\secsym\rightbracket%
\string\secno\the\secno\string\meqno\the\meqno}\immediate\closeout\lfile}}
\def\writestoppt{}\def\writedef#1{}
\def\seclab#1{\xdef #1{\the\secno}\writedef{#1\leftbracket#1}\wrlabeL{#1=#1}}
\def\subseclab#1{\xdef #1{\secsym\the\subsecno}%
\writedef{#1\leftbracket#1}\wrlabeL{#1=#1}}
\newwrite\tfile \def\writetoca#1{}
\def\leaderfill{\leaders\hbox to 1em{\hss.\hss}\hfill}
\def\writetoc{\immediate\openout\tfile=toc.tmp 
   \def\writetoca##1{{\edef\next{\write\tfile{\noindent ##1 
   \string\leaderfill {\noexpand\number\pageno} \par}}\next}}}

%
\catcode`\@=12 
%
\edef\tfontsize{\ifx\answ\bigans scaled\magstep2\else scaled\magstep4\fi}
 \tfontsize  \tfontsize
 \tfontsize \font\titlei=cmmi10 \tfontsize
\font\titleis=cmmi7 \tfontsize \font\titleiss=cmmi5 \tfontsize
\font\titlesy=cmsy10 \tfontsize \font\titlesys=cmsy7 \tfontsize
\font\titlesyss=cmsy5 \tfontsize  \tfontsize
 \tfontsize  \tfontsize
 \tfontsize 
\font\anbfsl=cmbxsl10 scaled\magstep1 \font\anbfsls=cmbxsl10
\font\anbf=cmbx10 scaled\magstep1
\skewchar\titlei='177 \skewchar\titleis='177 \skewchar\titleiss='177
\skewchar\titlesy='60 \skewchar\titlesys='60 \skewchar\titlesyss='60
 \ifx\answ\bigans\else scaled\magstep1\fi
\ifx\answ\bigans\else

 \font\absi=cmmi10 scaled\magstep1
\font\absis=cmmi7 scaled\magstep1 \font\absiss=cmmi5 scaled\magstep1
\font\abssy=cmsy10 scaled\magstep1 \font\abssys=cmsy7 scaled\magstep1
\font\abssyss=cmsy5 scaled\magstep1 
\skewchar\absi='177 \skewchar\absis='177 \skewchar\absiss='177
\skewchar\abssy='60 \skewchar\abssys='60 \skewchar\abssyss='60
\fi
\font\ninerm=cmr9 \font\sixrm=cmr6 \font\ninei=cmmi9 \font\sixi=cmmi6 
\font\ninesy=cmsy9 \font\sixsy=cmsy6 \font\ninebf=cmbx9 
\font\nineit=cmti9 \font\ninesl=cmsl9 \skewchar\ninei='177
\skewchar\sixi='177 \skewchar\ninesy='60 \skewchar\sixsy='60 
\def\ninepoint{\def\rm{\fam0\ninerm}
\textfont0=\ninerm \scriptfont0=\sixrm \scriptscriptfont0=\fiverm
\textfont1=\ninei \scriptfont1=\sixi \scriptscriptfont1=\fivei
\textfont2=\ninesy \scriptfont2=\sixsy \scriptscriptfont2=\fivesy
\textfont\itfam=\ninei \def\it{\fam\itfam\nineit}\def\sl{\fam\slfam\ninesl}%
\textfont\bffam=\ninebf \def\bf{\fam\bffam\ninebf}\rm} 
%
%

\hyphenation{anom-aly anom-alies coun-ter-term coun-ter-terms}
\def\inv{^{\raise.15ex\hbox{${\scriptscriptstyle -}$}\kern-.05em 1}}

\def\Dsl{\,\raise.15ex\hbox{/}\mkern-13.5mu D} 
\def\dsl{\raise.15ex\hbox{/}\kern-.57em\partial}

\def\lspace{\ifx\answ\bigans{}\else\qquad\fi}
\def\lbspace{\ifx\answ\bigans{}\else\hskip-.2in\fi} 
\def\boxeqn#1{\vcenter{\vbox{\hrule\hbox{\vrule\kern3pt\vbox{\kern3pt
	\hbox{${\displaystyle #1}$}\kern3pt}\kern3pt\vrule}\hrule}}}
\def\mbox#1#2{\vcenter{\hrule \hbox{\vrule height#2in
		\kern#1in \vrule} \hrule}}  
%

\def\darr#1{\raise1.5ex\hbox{$\leftrightarrow$}\mkern-16.5mu #1}

\def\roughly#1{\raise.3ex\hbox{$#1$\kern-.75em\lower1ex\hbox{$\sim$}}}

\lref\walba{D.M.~Walba and N.A.~Clark,
\underbar{Ferroelectrics}, \underbar{84}, 65 (1988).}

\lref\KN{See, for instance, 
R.D.~Kamien and D.R.~Nelson, Phys. Rev. E, \underbar{53}, 650 (1996);
\underbar{Phys. Rev. Lett.}, \underbar{74}, 2499 (1995).}

\lref\DGP{P.-G.~de Gennes and J.~Prost, \underbar{The Physics of Liquid Crystals},
Second
Edition, (Oxford University Press, New York, 1993).}

\lref\LL{L.D.~Landau and
E.M.~Lifshitz, \underbar{Quantum Mechanics}, Third Edition (Pergamon Press, Oxford,
1977).}

\input epsf
\header

\def\runauth{\hskip 0.17truecm RANDALL D. KAMIEN}
\def\runtitle{\hskip 0.17truecm CHIRAL INTERACTIONS AND STRUCTURES}

\def\up#1{\leavevmode \raise.16ex\hbox{#1}}

\def\bold#1{\setbox0=\hbox{$#1$}%
     \kern-.010em\copy0\kern-\wd0
     \kern.025em\copy0\kern-\wd0
     \kern-.020em\raise.0200em\box0 }

\lref\HKL{A.B.~Harris, R.D.~Kamien and T.C.~Lubensky,
\underbar{Phys. Rev. Lett.},
\underbar{78}, 1476 (1997); 2867 (1997).}

\lref\RMP{A.B.~Harris, R.D.~Kamien and T.C.~Lubensky,
\underbar{Rev. Mod. Phys.}, \underbar{71}, in press (1999).}

\def\oneskip{\vskip 0.166truein}
\settabs\+aaaaa&&\cr
\pageno=1
\voffset 0.8truecm
\hoffset 0.7truecm
{}~
\oneskip\oneskip\oneskip
{\+&CHIRAL INTERACTIONS AND STRUCTURES\cr}
\oneskip\oneskip\oneskip
{\+&RANDALL D. KAMIEN\cr}
\oneskip\oneskip
{
\baselineskip=0.16truein
{\+&Department of Physics and Astronomy, University of Pennsylvania,\cr}
{\+&Philadelphia, PA 19104, USA\cr}
\oneskip\oneskip\oneskip
}
\+&\hbox{\vbox{\baselineskip 0.16truein\hsize=13.1truecm
\noindent\underbar{Abstract}
 The chiral structure of liquid crystalline phases arises due
to the intrinsic chirality of the constituent mesogens.   While it
is seemingly straightforward to quantify the macroscopic chirality
by using, for instance, the cholesteric pitch or the optical
rotatory power, it is not as simple to quantify the chirality of
a single molecule.  I will discuss a systematic quantification of
molecular chirality and show how the resulting chiral parameters
may be used to predict macroscopic chiral structure.
}}\cr
\oneskip
\ifx\preflag\prepr\centerline{(19 August 1999)}\else\oneskip\fi\oneskip
\noindent\underbar{SUMMARY}\oneskip

Liquid crystal physics is a useful and fruitful proving ground for ideas based
on symmetry.  The entire apparatus and formalism of spontaneous symmetry
breaking can be used to accurately model and predict behavior
at phase transitions between the host of liquid crystalline phases as well
as well as inside the phases themselves\KN .  In addition, many of the
phenomenological parameters can be estimated rather well through dimensional
analysis.
To illustrate, we consider the nematic phase.
The nematic is a phase with a spontaneously chosen direction, which we can take
to
be the $\hat z$-axis.  There are two Goldstone modes associated with
infinitesimal
rotations away from this preferred axis, $\delta n_x$ and $\delta n_y$, the
deviation of the unit nematic director ${\bf n}$ away from ${\bf n}_0=\hat z$.
The free
energy dictated by the nematic symmetry (${\bf n}\rightarrow -{\bf n}$) and
rotational invariance is the Frank free energy\DGP:
\eqn\efrank{F={1\over 2}\int d^3\!x\,\left\{K_1\left(\nabla\cdot{\bf
n}\right)^2
+ K_2\left[{\bf n}\cdot\left(\nabla\times{\bf
n}\right)\right]^2 +K_3\left[{\bf n}\times\left(\nabla\times{\bf
n}\right)\right]^2\right\}}

The elastic constants have the dimensions of energy per unit length.  Since
typical nematic phases exist within an order-of-magnitude of room temperature,
the relevant energy scale is $k_{\scriptscriptstyle B}T\sim 4\times
10^{-14}\;{\rm
erg}$. The lengthscale is simply the molecular size, roughly $10\rm\AA$.
Put together, we can estimate the three Frank constants, $K_1$, $K_2$ and $K_3$
to be on the order of $1$ microdyne.  Indeed, this simple estimate is
remarkably
good for typical nematics\DGP .  When the constituent molecules are chiral,
a new term is allowed in \efrank\ which is not invariant under inversion:
\eqn\efrankc{\delta F^* = \int d^3\!x\,K_2q_0\left[{\bf
n}\cdot\left(\nabla\times{\bf
n}\right)\right]}
where the new parameter $q_0$ is a pseudoscalar, reflecting the chirality
of the molecules.  However, it is clear that this new parameter introduces
a {\sl lengthscale}, $q_0^{-1}$.  Indeed, though the ground state of \efrank\
is simply
a constant director field, the ground state of
$F + \delta F^*$ is not nematic, but {\sl cholesteric}, with
${\bf n}=\left[\cos(q_0 z),\sin(q_0 z), 0\right]$, corresponding to
a helical structure with a pitch $P=2\pi/q_0$ which is typically
on the order of microns.  Thus the presence of chirality
introduces a {\sl new} lengthscale into the problem which leads to a number of
macroscopic, periodic structures, of which the cholesteric is only one example.
It is a theoretical challenge to predict and explain these
new lengthscales.  While the elastic constants of the
cholesteric can be estimated on dimensional grounds, the micron-sized pitch
is not comparable to {\sl any} molecular scale.

We have
found a systematic approach to calculate
macroscopic chiral properties from the microscopic character of the
constituents.  A chief result is an expression for the cholesteric
wavenumber $q_0$.  While dimensional analysis predicts something too large,
we have shown\HKL\ that any consistent mean-field treatment predicts $q_0=0$.  This
is a straightforward consequence of group theory:  in a uniaxial
phase, there is no long range ordering of the molecular orientation
{\sl perpendicular} to the nematic director.  It follows that the average
electron and nucleon density of the molecules has a ${\cal C}_\infty$ axis
along the director.  However, there are no symmetry groups that have
a ${\cal C}_\infty$ axis which do not contain the inversion element\LL.  Thus
the uniaxial averaging ``washes out'' the chirality of the molecules and
hence $q_0$ must vanish.

\centerline{\vbox{\epsfxsize=4truein\hsize=4truein\epsfbox{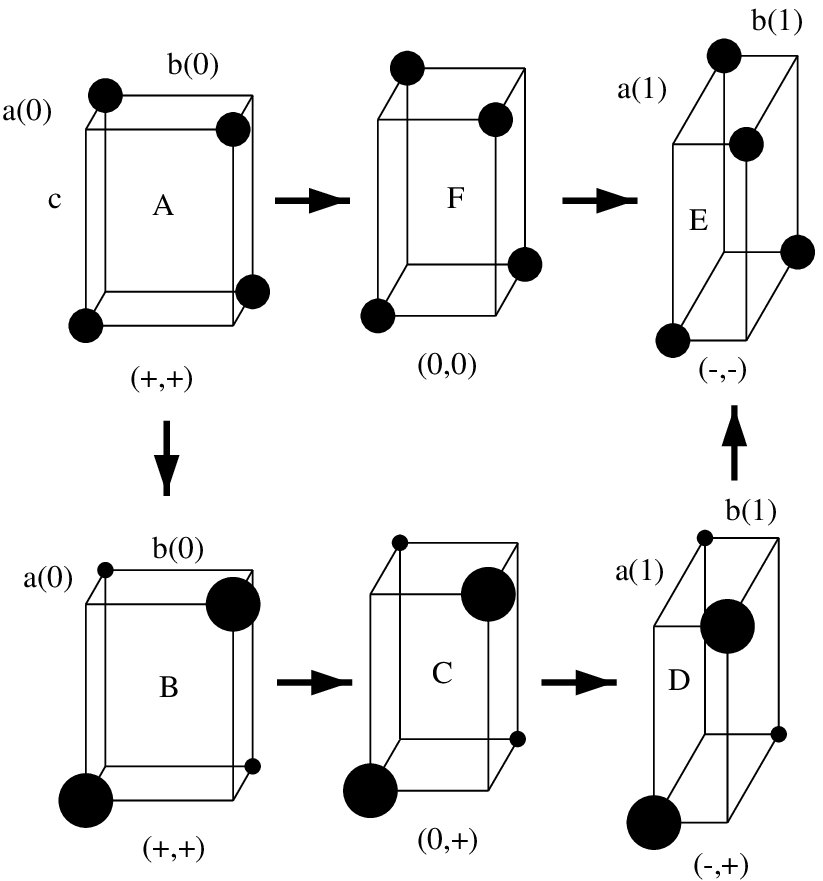}}}\vskip
0.25truein
\centerline{\vbox{\baselineskip=0.20truein\hsize=4truein
\noindent FIGURE 1. Path I ($\bf AFE$) and path II
($\bf ABCDE$)
between
chiral enantiomers
(mirror images) $\bf A$ and $\bf E$ for a molecule consisting of $4$ atoms at
the tetrahedral vertices of a cube.
Below each configuration we give
$(\sigma, \sigma')$, where $\sigma$ and $\sigma'$ are
respectively the signs or zero value of the chiral parameters $\psi_0$ and
$\psi_0'$. (From reference 5).
}}
We have argued that it is therefore necessary to include correlations between
the orientations of the molecules and found that\HKL :
\eqn\eresult{
K_2 q_0 = \psi \left\langle\,g_B(R) V(R)\,\right\rangle}
where $g_B(R)$ is the biaxial-orientation correlation function, $V(R)$ is
a function of the interatomic potentials, $\langle\,\cdot\,\rangle$ in
\eresult\ is the fluid average of the intermolecular separation $R$ and
$\psi$ is a chiral parameter depending only on a single molecule.  Thus, if
the biaxial correlations are absent $q_0=0$.

It is interesting that this result also depends on a pseudoscalar molecular
parameter
$\psi$.  Being a pseudoscalar, $\psi\rightarrow -\psi$ under inversion.  Thus
this parameter can only be non-zero for chiral molecules.
However, this parameter also {\sl quantifies} \hfill\break
\centerline{\vbox{\epsfxsize=4truein\hsize=4truein\epsfbox{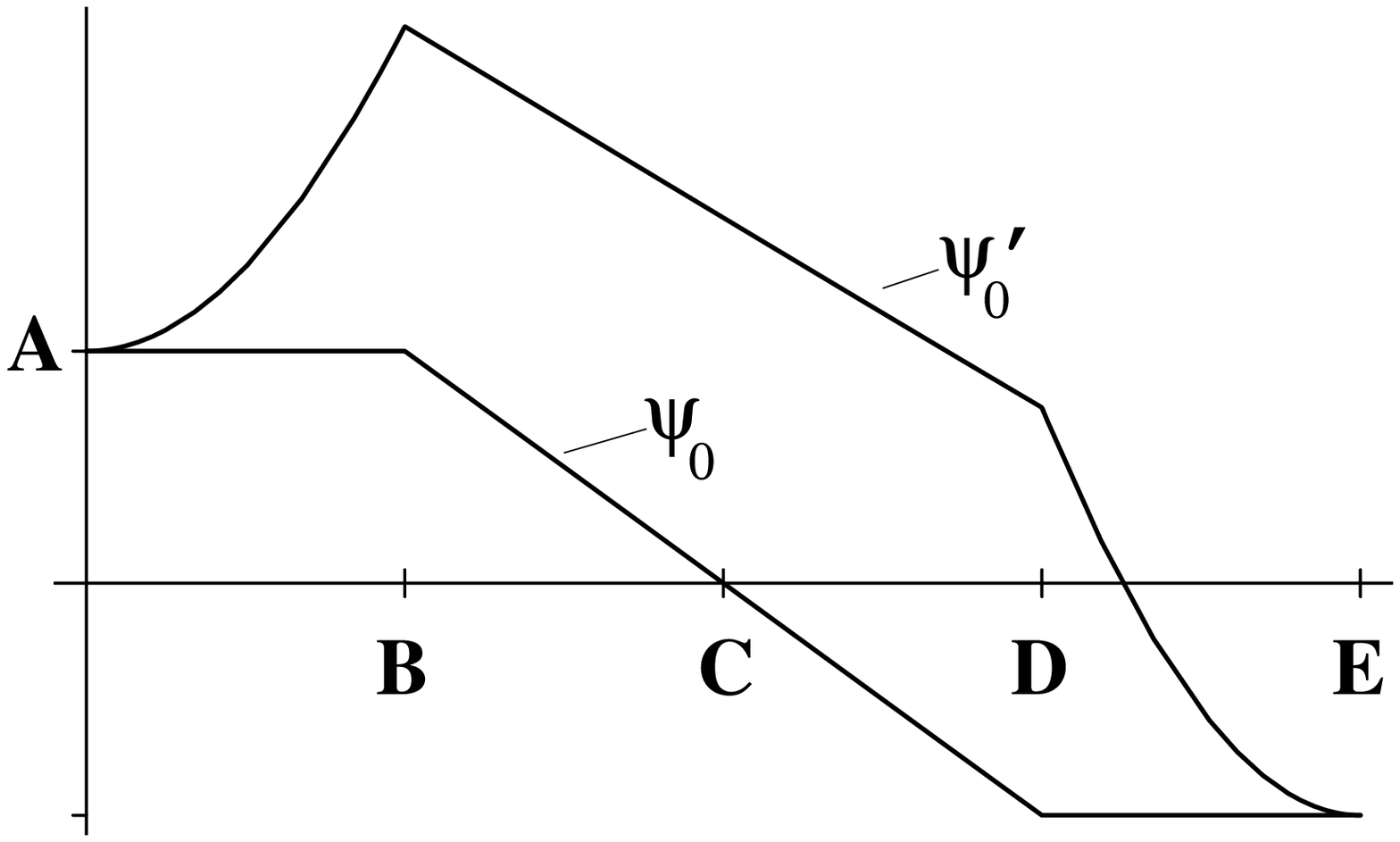}}}\vskip
0.25truein
\centerline{\vbox{\baselineskip=0.20truein\hsize=4truein
\noindent FIGURE 2. Plot of two chiral measures
$\psi_0$ and
$\psi'_0$ along the path
${\bf ABCDE}$.  Note that both parameters pass through zero on this path, but
they do not pass through zero at the same place.  On the
Path ${\bf AFE}$ both parameters vanish simultaneously. (From reference 5)
}}

\smallskip

\noindent the chirality at a molecular
level and
makes {\sl a direct connection} between the microscopics and the macroscopic
parameter $q_0$. However, when
calculating other macroscopic quantities other than $q_0$ (or when calculating
higher-order corrections to $q_0$) other chiral parameters appear which
are not equal to $\psi$.  In fact, they may not even have the same sign
as $\psi$!  To make this concrete, we have considered ``rubber glove
molecules''
which can be transformed into their enantiomer through a chiral pathway \RMP .
Figure
1 illustrates the chiral pathway and Figure 2 shows the development of
two different chiral parameters $\psi_0$ and $\psi_0'$.  Notice that
they vanish at different places and thus need not have the same sign
over the entire range of deformation of the molecule.
Even
though ``left'' and ``right'' are a matter of convention, our construction
shows that even a chosen convention cannot be defined globally since 
continuous deformations can change ``left'' to ``right'' without
an achiral intermediate state.  
Thus the handedness of an object depends on the property of interest.  
\bigskip
\noindent\underbar{ACKNOWLEDGEMENTS}\oneskip
It is a pleasure to acknowledge stimulating discussions with my collaborators,
A.B.~Harris and T.C.~Lubensky, as well
as with S.~Fraden,
R.~Meyer, D.~Nelson,
V.A.~Parsegian, R.~Pelcovits, R.~Podgornik, H.~Strey and D.~Walba.
This work was supported by NSF Career Award DMR97-32963, an award from
Research Corporation, a gift from L.J. Bernstein
and the Alfred P. Sloan Foundation.

\footatend\bigskip\immediate\closeout\rfile\writestoppt
\baselineskip=.20truein\leftline{\underbar{REFERENCES}}\bigskip{\frenchspacing%
\parindent=20pt\escapechar=` \input refs.tmp\vfill\eject}\nonfrenchspacing

\bye